\documentclass[preprint,12pt]{aastex}
\usepackage{amssymb}
\usepackage{natbib}
\usepackage{lscape}
\begin{document}

\title{THE FAINT END OF THE QSO LUMINOSITY FUNCTION AT $z=3$\altaffilmark{1}}
\author{\sc Matthew P. Hunt, Charles C. Steidel\altaffilmark{2}}
\affil{Palomar Observatory, California Institute of Technology,
	Department of Astronomy, MS 105-24, Pasadena, CA 91125}
\author{\sc Kurt L. Adelberger}
\affil{Carnegie Observatories, 813 Santa Barbara Street, Pasadena, CA 91101}
\and
\author{\sc Alice E. Shapley}
\affil{Department of Astronomy, University of California,
	Berkeley, CA 94720}

\email{mph@astro.caltech.edu}

\slugcomment{DRAFT: \today}

\altaffiltext{1}{Based, in part, on data obtained at the 
W.M. Keck Observatory, which is operated as a scientific partnership
among the California Institute of Technology, the University of
California, and NASA, and was made possible by the generous financial
support of the W.M. Keck Foundation.}  
\altaffiltext{2}{Packard Fellow}

\begin{abstract}
  We present the first measurement of the faint end of the QSO
  luminosity function at $z=3$.  The QSOs, which range from
  $M_{1450}=-21$ to $M_{1450}=-27$, were discovered in 17 fields
  totaling $0.43~{\rm deg^2}$ using multicolor selection criteria (the
  Lyman break technique) and spectroscopic followup.  We find that the
  faint-end slope of the luminosity function is $\beta_l=1.24 \pm
  0.07$ ($\Phi \propto L^{-\beta_l}$), flatter than the value of
  $\beta_l=1.64 \pm 0.18$ measured at lower redshift.  The integrated
  rest 1450\AA\ UV luminosity of $z=3$ QSOs is only 50\% of most
  previous estimates, and is only $\simeq 8$\% of that produced by
  Lyman break galaxies at the same redshifts.  Assuming that ionizing
  photons from faint QSOs are as successful in escaping their host
  galaxies as bright QSOs, we estimate the total contribution of QSOs
  to the ionizing flux $J_{912}$ at $z \sim 3$, $J_{912} \simeq 2.4
  \times 10^{-22}$ ergs s${-1}$ cm$^{-2}$ Hz$^{-1}$.  This estimate,
  which we regard as an upper limit, remains consistent with rough
  estimates of $J_{912}$ based on the Lyman $\alpha$ forest
  ``proximity effect.''
\end{abstract}

\section{Introduction}
\label{sec:intro}

The QSO luminosity function (LF) at high redshifts provides important
constraints on the ionizing UV radiation field of the early universe.
Until now, however, the faint end of the QSO LF has not been measured
at high redshift.  Instead, low-redshift measurements of the faint end
were combined with high-redshift measurements of the bright end to
estimate the entire LF at high redshift.  Various models of LF
evolution have been proposed; for example, a model proposed by
\citet{pei95} consists of a double power law~\citep{bsp88} whose
bright- and faint-end slopes are independent of redshift, and whose
power-law break $L_z(z)$ comes at a luminosity which is proportional
to a Gaussian in $z$, with a maximum near $z_\star=2.75$ and $\sigma =
0.93$ redshift.  This model is representative of ``pure luminosity
evolution'' models, as the overall normalization and the power-law
slopes are independent of redshift.  While pure luminosity evolution
has been shown to work well at $z < 2.3$ \citep{bsp88,2dF-lf}, there
is now evidence that it is insufficient at high redshift; for example,
SDSS results demonstrate that the bright-end slope is flatter at $z >
3.6$ than in the local universe~\citep{fan01}.  The luminosity of the
power law break and the faint-end slope have not been measured at high
redshifts prior to the survey presented here.

\begin{figure}
  \plotone{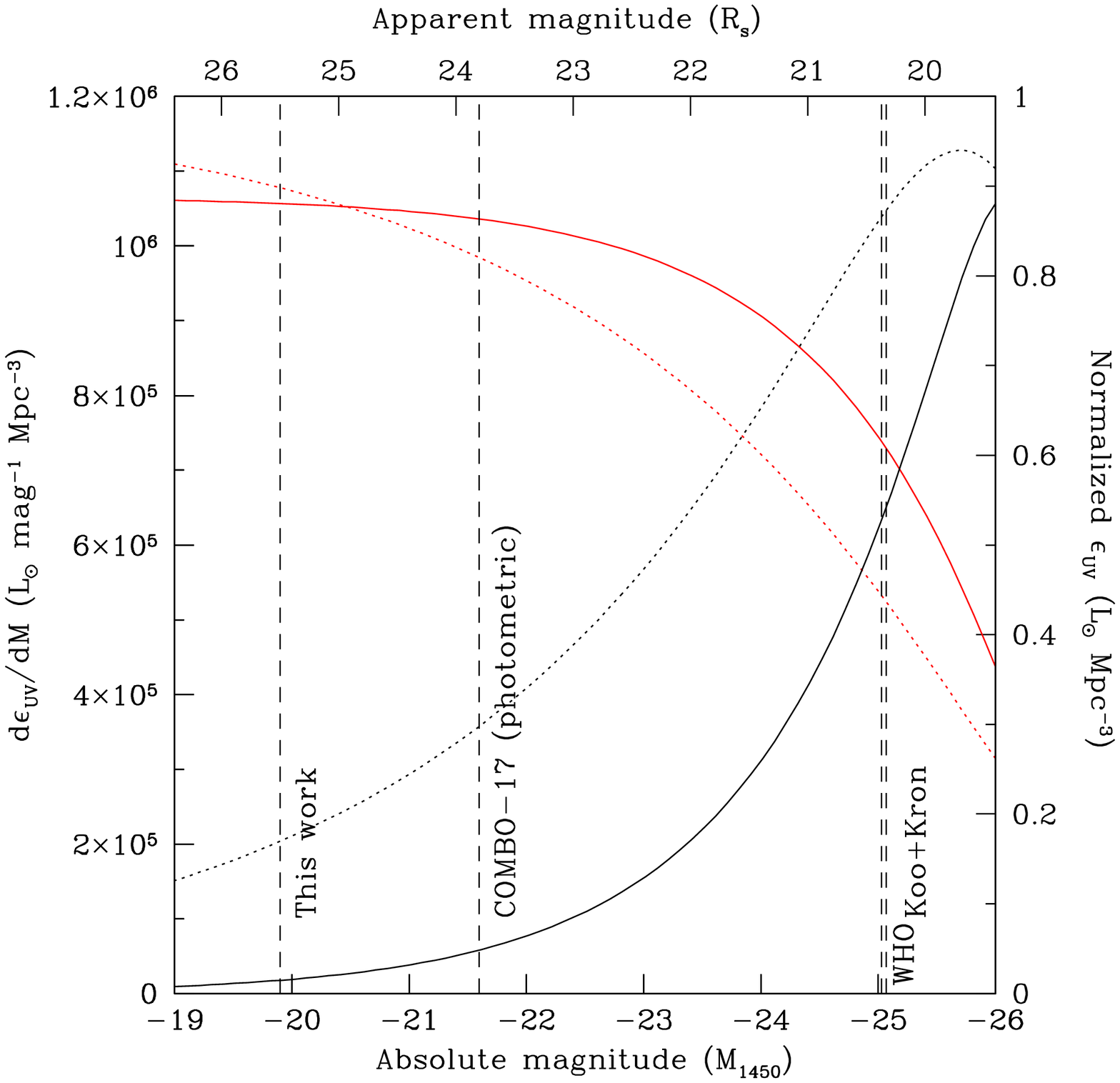} \caption{The $1450~{\rm\AA}$ UV luminosity
  density produced by QSOs as a function of absolute magnitude,
  assuming the $z=3$ luminosity function of \citet{pei95} (dashed) and
  our fit described in Section~\ref{sec:lf} (solid).  The differential
  luminosity density is shown by the black curve and is read from the
  left scale.  The cumulative luminosity density is shown by the red
  curve and read from the right scale, which is normalized to the
  total.  The magnitude limits of this work and other surveys are
  indicated by dashed lines.  The corresponding apparent ${\cal R}_s$
  magnitude is indicated at the top.  It is immediately clear that
  this work explores a substantially fainter portion of the luminosity
  function and accounts for virtually all of the total UV luminosity
  from QSOs. } \label{fig:density}
\end{figure}

We have made the first direct measurement of the faint end of the
QSO~LF at high redshift, using a sample of 11 faint $z \sim 3$ QSOs
discovered in a survey for Lyman break galaxies.
Figure~\ref{fig:density} illustrates the depth of this survey relative
to previous $z=3$ QSO surveys and demonstrates that the vast majority
of the total QSO UV luminosity arises from QSOs bright enough to be
included in this survey.

Throughout this paper, the term ``QSO'' is used to describe all
broad-lined AGN without imposing the traditional $M_B < -23$
luminosity cutoff.  Spectral properties of such objects are
essentially the same across at least two decades of
luminosity~\citep{ccs-agn}, thus we find no reason to impose such a
cutoff.  In Section~\ref{sec:survey}, we will present an overview of
the survey parameters and photometric criteria for candidate
selection.  In Section~\ref{sec:completeness}, we will describe our
measurements of photometric and spectroscopic completeness, and our
calculation of the survey effective volume.  The QSO luminosity
function will be presented in Section~\ref{sec:lf}, followed by a
discussion of its implications for the UV radiation field in
Section~\ref{sec:implications}.

\section{Survey information}
\label{sec:survey}

The Lyman break technique has proved to be a successful and efficient
means of photometrically identifying star-forming galaxies and AGN at
$z=3$~\citep{ccs-lbg}.  Similar multicolor approaches have been used
in previous, shallower surveys for high-redshift QSOs with good
success~\citep[e.g.][]{kkc86, who-1, poss}.  Survey fields were imaged
in $U_n$ (effective wavelength $3550~\textrm\AA$), $G$
($4730~\textrm\AA$), and ${\cal R}$ ($6830~\textrm\AA$)
filters~\citep{steidelhamilton93}.  A star-forming galaxy at $z=3$
will have a Lyman break in its SED that falls between the $U_n$ and
$G$ filters, resulting in a $U_n-G$ color that is substantially redder
than its $G-{\cal R}$ color.  Objects meeting the following
photometric criteria were selected as candidate $z=3$ galaxies:
\begin{eqnarray}
  {\cal R} & > & 19\label{eq:photfirst}\\
  {\cal R} & < & 25.5\\
  G-{\cal R} &<& 1.2\\
  G-{\cal R}+1.0 &<& U_n-G
\label{eq:photlast}
\end{eqnarray}
At $z=3$, the intergalactic medium provides sufficient opacity to also
select many QSOs with this technique even if their intrinsic SED lacks
a strong Lyman break.  The details are addressed in
Section~\ref{sec:completeness}.

The LBG survey fields used for this study cover 0.43~${\rm deg}^2$ in
17~fields, which are discussed in detail in~\citet{ccs-lbg}.  A
composite spectrum and other information relating to the 13 QSOs
discovered in the survey have already been published~\citep{ccs-agn}.
Two of these QSOs satisfied earlier versions of the photometric
criteria, but do not satisfy the final versions listed above, and have
been excluded from this paper's results.  The sample discussed in this
paper, therefore, comprises 11~QSOs.

\section {Sensitivity to QSOs}
\label{sec:completeness}

\subsection{Photometric completeness}
\label{ss:photometric}

The intrinsic $U_n-G$ and $G-{\cal R}$ colors of QSOs depend primarily
on the spectral index of their continuum and their
Lyman-$\alpha$+\ion{N}{5} equivalent width.  To measure the
distribution of intrinsic colors (i.e. without the effects of
measurement error), we produced a template QSO spectrum consisting of
59 QSOs studied by~\citet[hereafter SSB]{ssb89}.  These QSOs were
discovered using objective prism techniques and are not expected to
have significant selection biases in common with multicolor selection
techniques.  The SSB QSOs are about 100 times brighter than LBG survey
QSOs, but a comparison of the SSB and LBG composite spectra suggests
that the two populations are sufficiently similar that using the SSB
composite as a template is satisfactory~\citep{ccs-agn}. An average
intergalactic absorption spectrum was used to absorption-correct the
template using the model of~\citet{madau95}, and portions of the
spectrum having poor signal-to-noise were replaced with a power-law
fit to the continuum.

\begin{deluxetable}{lrr}
  \tablewidth{0pt}
  \tablecaption{The mean and sigma of the Gaussian distributions
    used for simulating the colors of QSOs (see
    section~\ref{ss:photometric}).  The \ion{C}{4} equivalent width was
    scaled in proportion to that of Ly$\alpha$+\ion{N}{5} in order to
    maintain the template's original line ratio.}
  \tablehead{
    \colhead{Parameter} &
    \colhead{Mean} &
    \colhead{Sigma}
    }
  \startdata
  Continuum slope ($F_\nu$) & 0.46 & 0.30\\
  EW(Ly$\alpha$+\ion{N}{5}) (\AA{}) & 80.0 & 20.0 
  \enddata
  \label{tab:distrib}
\end{deluxetable}

The template spectrum was repeatedly altered to have continuum slopes
and Lyman-$\alpha$ equivalent widths drawn from the Gaussian
distributions described in Table~\ref{tab:distrib}, a compromise
between the results of~\citet{vandenberk01},~\citet{fan3}, and our SSB
template. Each altered spectrum was redshifted to 40 redshifts
spanning $z=2.0$ to $z=4.0$.  Intergalactic absorption was added by
simulating a random line-of-sight to each QSO with absorbers
distributed according to the MC-NH model of~\citet{bershady99}.  (For
comparison, an average intergalactic extinction curve \citep{madau95}
was also employed.  The results were not significantly different.)
The spectrum was then multiplied by our filter passbands to produce a
distribution of intrinsic colors which reflects the QSO population.

\begin{figure}
  \plotone{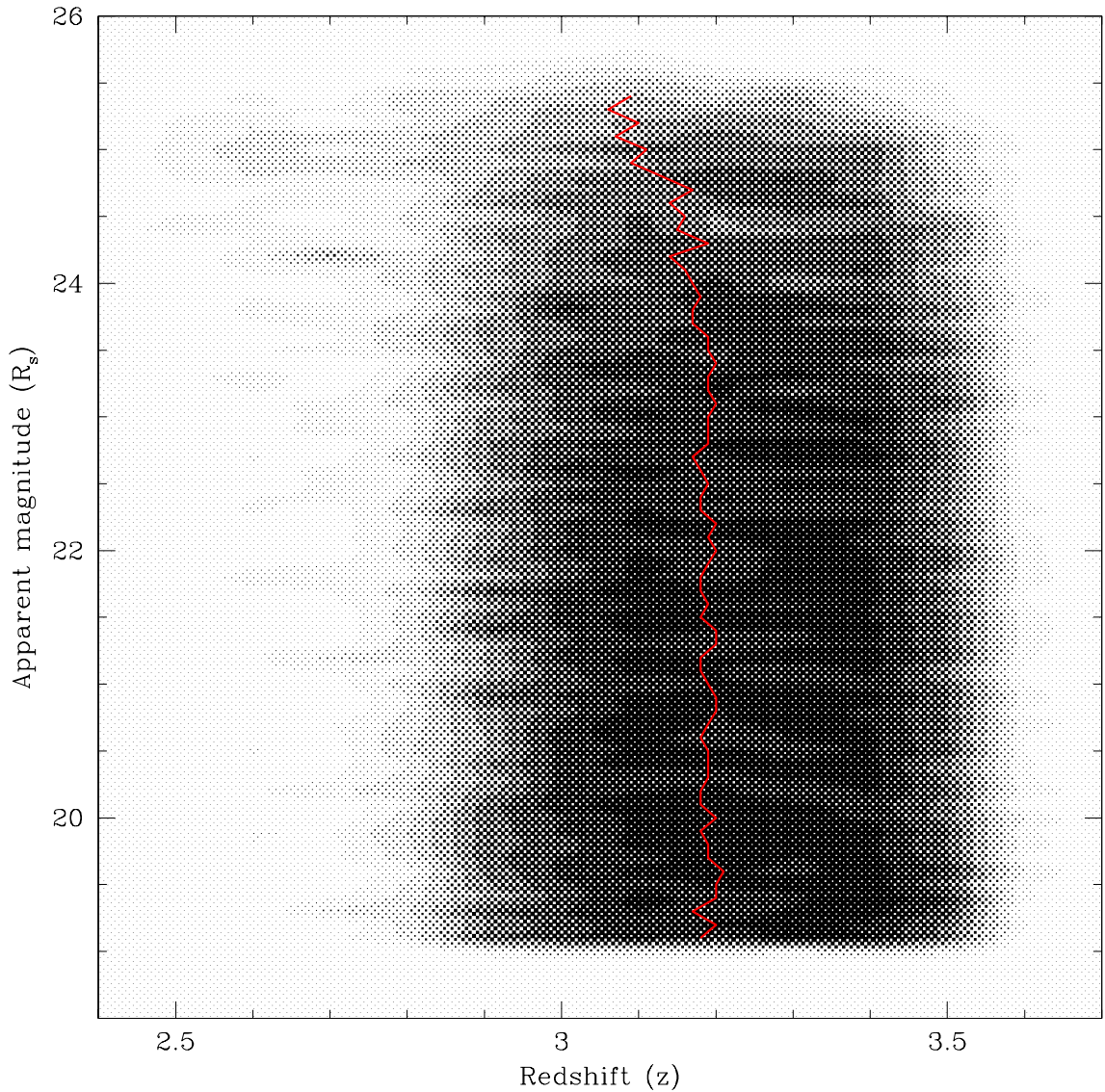} \caption{The fraction of simulated QSOs having
  measured colors that satisfy the photometric selection criteria
  (equations~\ref{eq:photfirst}--\ref{eq:photlast}) as a function of
  redshift and apparent ${\cal R}_s$ magnitude.  The QSOs were
  simulated using the method described in
  Section~\ref{ss:photometric}.  This plot does not include the
  effects of spectroscopic incompleteness.  The centroid of the
  distribution, as a function of magnitude, is marked with a red
  curve.  The grayscale levels are evenly spaced at 10\% intervals;
  the darkest level represents completeness in excess of 90\%.}
  \label{fig:fphot}
\end{figure}

These colors were used to place artificial QSOs into the survey
images.  5000~QSOs drawn uniformly from the redshift interval $2.0 < z
< 4.0$ and apparent magnitude interval $18.5 < {\cal R} < 26$ were
simulated in each of the 17 survey fields.  The apparent magnitude
interval is 0.5 magnitudes larger than the selection window on each
end, in order to allow measurement errors to scatter objects into the
selection window.  The artificial QSOs added to an image were given
radial profiles matching the PSF of that image (i.e. they were assumed
to be point sources).  This assumption has little practical effect,
because even galaxies are barely resolved at $z=3$, and no
morphological criteria were applied to candidates during the LBG
survey.  The images were processed using the same modified
FOCAS~\citep{focas} software which was used for the actual candidate
selection, and the observed colors of the simulated objects were
recorded.  The intrinsic QSO color distribution was thus transformed
to an observed color distribution.  Figure~\ref{fig:fphot} shows the
fraction of simulated QSOs that meet the photometric selection
criteria as a function of redshift.  As a consistency check, the curve
shown was multiplied by $\int \Phi(L,z)~dL$ to reflect the underlying
redshift distribution of QSOs, and compared to the distribution of
QSOs discovered in this survey, using a Kolmogorov-Smirnov test.  The
result was $P=0.49$ using the \citet{pei95} LF, and $P=0.45$ using the
LF shape measured in Section~\ref{sec:lf}, indicating consistency
between the expected and actual distribution of QSO redshifts.

\subsection{Spectroscopic completeness}
\label{ss:spectroscopic}

With the observed color distribution, we can measure the fraction of
QSOs which meet the LBG color criteria as a function of absolute
magnitude and redshift.  In order to determine the effective volume of
the survey, it is also necessary to know the probability of a
photometric candidate being observed spectroscopically.  At faint
apparent magnitudes (${\cal R} > 23$), there were 2,289 candidates in
the 17 fields, enough to measure the spectroscopic observation
probability as a function of (${\cal R}$, $G-{\cal R}$).  The
photometric candidates were divided into bins in (${\cal R}$, $G-{\cal
R}$) parameter space, using an adaptive bin size which increases
resolution where the parameter space is densely filled with
candidates.  The probability of spectroscopic observation was measured
for each bin.

At ${\cal R} < 23$, there are too few photometric candidates to obtain
an accurate measurement of the selection probability.  However, at
these apparent magnitudes, candidates with relatively blue $G-{\cal
  R}$ were likely to be QSOs and hence were nearly always observed
spectroscopically.  Candidates with red $G-{\cal R}$ were likely to be
stellar contaminants, and were less likely to be observed.  Hence we
have estimated the probability of spectroscopic observation to be
unity for candidates with observed magnitudes ${\cal R} < 23$ and
$G-{\cal R} < 1$, and 0.5 for candidates with ${\cal R} < 23$ and
$G-{\cal R} > 1$.  The results are insensitive to the latter value
because the observed colors of QSOs are rarely observed to be so red.

We assume that any spectroscopically observed QSO will be identified
as such and a redshift obtained, since our spectroscopic integration
times were chosen so that we could often identify faint LBGs using
their absorption lines (typically 90 minutes using Keck--LRIS).
Because QSOs have strong, distinctive emission lines they are easily
identifiable even at the faintest apparent magnitudes in the survey
(${\cal R} = 25.5$).

\subsection{Effective volume of the survey}
\label{ss:veff}

For comparison with other work, e.g. SDSS, we wish to measure the QSO
luminosity function with respect to 1450~\AA{} rest frame AB absolute
magnitude ($M_{1450}$).  At any given redshift in this study, we
estimate an object's apparent magnitude $m_{1450}$ as a linear
combination of its ${\cal R}$ and $G$ magnitudes.  A small
redshift-dependent correction, derived from our simulations of QSO
colors, was then made to the value.  This correction, typically of
order $0.25~{\rm mag}$, accounts for Ly-$\alpha$ emission in $G$,
Ly-$\alpha$ forest absorption, and similar effects.  If we denote by
$f_{\rm phot}(m_{1450},z)$ the probability that a QSO of apparent
1450~\AA{} rest frame AB magnitude $m_{1450}$ and redshift $z$ will
have observed colors and magnitudes that meet the selection criteria
for LBGs, and we denote by $f_{\rm spec}(m_{1450},z)$ the fraction of
such candidates that will be observed spectroscopically, we can
measure the effective volume of the survey as a function of absolute
magnitude,
\begin{equation}
  V_{\rm eff}(M) = \int_{\Omega}
  \int_{z=0}^{z=\infty}~f_{\rm phot}(m_{1450}(M,z),z)~f_{\rm
    spec}(m_{1450}(M,z),z)~\frac{dV}{dz~d\Omega}~dz~d\Omega
\end{equation}
where $m_{1450}(M_{1450},z)$ is the apparent magnitude corresponding
to absolute magnitude $M_{1450}({\cal R},G,z)$, $\Omega$ is the solid
angle of the survey, and $dV/dz~d\Omega$ is the co-moving volume
element corresponding to a redshift interval $dz$ and solid angle
$d\Omega$ at a redshift $z$ and using an assumed cosmology.  This
approach is explained in detail by \citet{adelberger2002}.  For this
measurement of the LF, we averaged $V_{\rm eff}$ over bins 1~mag in
width.

\section{The luminosity function}
\label{sec:lf}

\begin{figure}
  \plotone{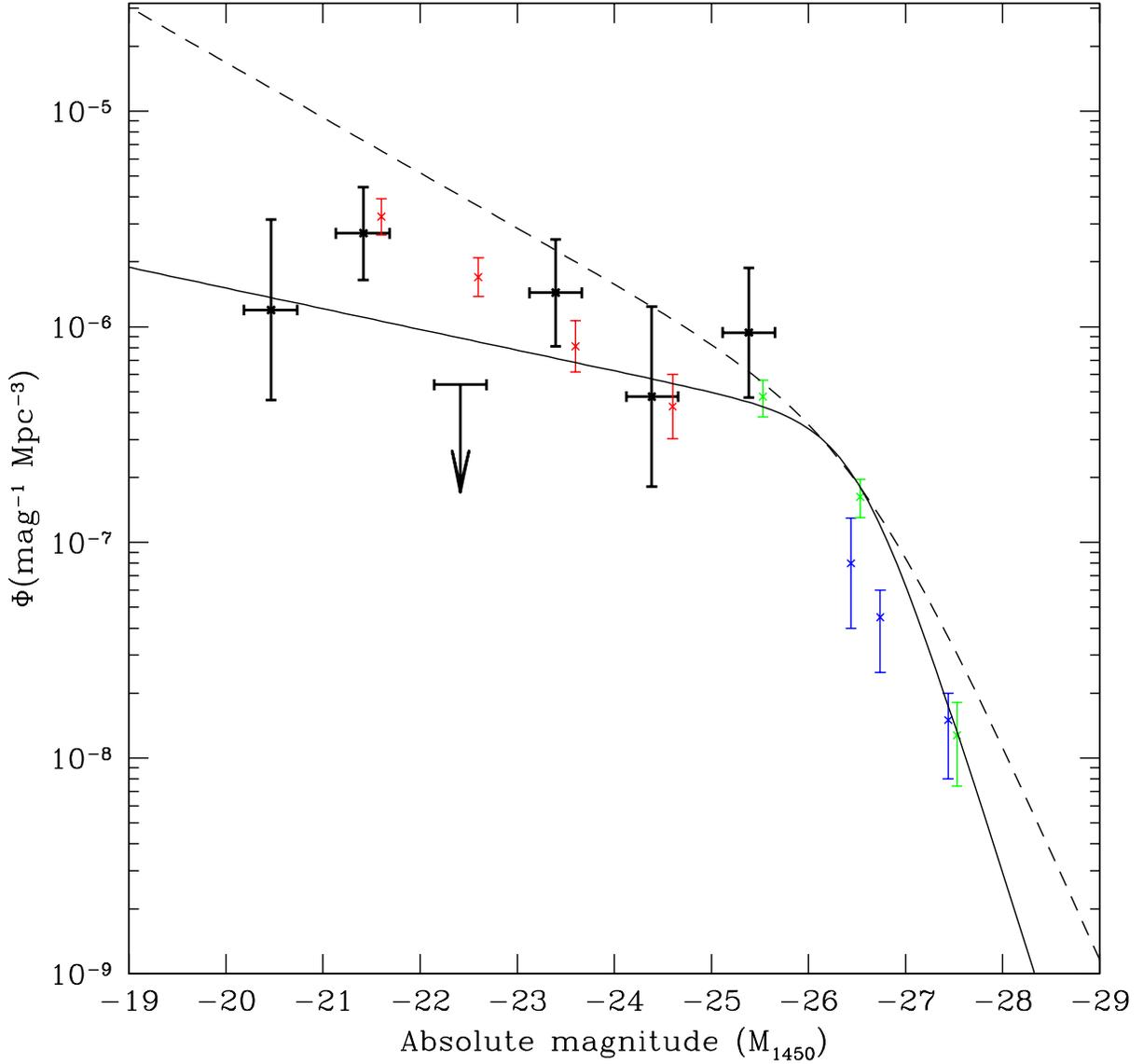} \caption{The faint end of the $z=3$ QSO luminosity
  function under an assumed $\Omega_m=1$, $\Omega_\Lambda=0$, $h=0.5$
  cosmology.  The vertical errorbars indicate 1-sigma uncertainties
  arising from Poisson statistics.  The horizontal errorbars indicate
  the expected rms scatter of magnitudes of detectable QSOs in each
  magnitude bin. The arrow indicates a 1-$\sigma$ upper limit from an
  empty bin.  The double power-law $z=3$ LF of \citet{pei95} has been
  converted from $M_B$ to $M_{1450}$ (using an assumed $\alpha=-0.5$
  continuum slope) and has been plotted as well using a dashed curve,
  and our fit is plotted with a solid curve.  SDSS
  points~\citep{fan01} from $3.6 < z < 3.9$ have been evolved to
  $z=3.0$ using the luminosity evolution of~\citet{pei95} and plotted
  in blue for comparison.  The $2.2 < z < 3.0$ and $3.0 < z < 3.5$
  points of WHO~\citep{who} have been combined and plotted in green,
  and the $2.4 < z < 3.0$ and $3.0 < z < 3.6$ points of
  COMBO-17~\citep{wolf03} have been combined and plotted in red.}
  \label{fig:lf}
\end{figure}

Having measured the effective volume of the survey as a function of
absolute magnitude, we can place points on the QSO luminosity function
simply by placing the observed QSOs in absolute magnitude bins and
dividing by the effective volume of the survey at that absolute
magintude.  A plot of the luminosity function is shown in
Figure~\ref{fig:lf}.  The vertical errorbars indicate 1-sigma
confidence intervals reflecting the Poisson statistics due to the
number of QSOs in the bin.  The uncertainty in $V_{\rm eff}$ is not
reflected, as the Poisson statistics dominate (e.g. there is only 1
QSO in the faintest bin, where imprecisions in photometry lead to the
greatest $V_{\rm eff}$ uncertainty).  The horizontal errorbars
indicate the rms width of the bin, weighted according to the effective
volume and expected luminosity function~\citep{pei95} as a function of
absolute magnitude; for a bin centered on $M=M_0$ and 1 magnitude
wide, the position of the point and its errorbar width are given by
\begin{eqnarray}
\langle M \rangle &=& \frac{\int_{M_0-1/2}^{M_0+1/2}
  M~V_{\rm eff}(M)~\Phi(M)~dM}{\int_{M_0-1/2}^{M_0+1/2}
  V_{\rm eff}(M)~\Phi(M)~dM} \\
\sigma_M &=& \left(\frac{\int_{M_0-1/2}^{M_0+1/2} (M-\langle M\rangle)^2
    ~V_{\rm eff}(M)~\Phi(M)~dM}
  {\int_{M_0-1/2}^{M_0+1/2} V_{\rm eff}(M)~\Phi(M)~dM}\right)^{1/2}
\end{eqnarray}
where $\Phi(M)$ is the $z=3$ luminosity function of \citet{pei95}; as
both the Pei LF and the observed points are quite flat at these
magnitudes, this calculation is insensitive to the precise slope
assumed for $\Phi(M)$, and retroactively trying our fitted value has
no significant effect on the results.  An $\Omega_m=1$,
$\Omega_\Lambda=0$, $h=0.5$ cosmology has been assumed for comparison
with previous work.

Comparison with the other points shown in Figure~\ref{fig:lf} suggests
that our results are largely consistent with previous
measurements~\citep{wolf03,who,fan01} in the region of overlap.  The
shape of the LF near the power law break is somewhat unclear, and our
present sample is unable to resolve this issue.  The total luminosity
of the LF is quite sensitive to the location of the break.  A
shallower, wide-field survey using identical LBG techniques is nearing
completion~\citep{hunt04}, and should better constrain the $-27 <
M_{1450} < -24$ portion of the luminosity function.

The observed faint-end slope appears to be considerably flatter than
$\beta_l = -1.64$ used by \citet{pei95}.  The Pei $z=3$ LF is shown as
the solid curve in Figure~\ref{fig:lf}.  In order to quantify the
difference, we have fit the double power law of \citet{bsp88},
identical in form to that used by Pei,

\begin{equation}
\Phi(L,z) = \frac{\Phi_\star/L_z}{(L/L_z)^{\beta_l} + (L/L_z)^{\beta_h}},
\end{equation}

where $\beta_l$ and $\beta_h$ are the faint- and bright-end slopes,
respectively, $L_z(z)$ is the luminosity of the power-law break, and
$\Phi_\star$ is the normalization factor.  We have combined our data
with those of \citet{who} to fit the entire luminosity function.  The
SDSS data plotted in Figure~\ref{fig:lf} were excluded because they
were measured at $z > 3.6$, and the authors have demonstrated redshift
evolution in the bright-end slope.  Given the relatively small number
of data points and large errorbars, fits for the four parameters
$(L_z, \Phi_\star, \beta_l, \beta_h)$ are degenerate.  We have
therefore assumed that the \citet{pei95} luminosity evolution model
still holds, and adopted the same values for $L_z$ and $\Phi_\star$.
A weighted least-squares fit for $\beta_l$ and $\beta_h$ was
performed, and the measured faint-end slope was $\beta_l=1.24 \pm
0.07$.  The measured bright-end slope was $\beta_h=4.56 \pm 0.51$, but
in addition to the large uncertainty, this parameter is highly
degenerate with the assumed parameters $L_z$ and $\Phi_\star$, and is
very sensitive to the brightest data point.  Likewise, the errorbars
for the faint-end slope are smaller than they would be for a general
four-parameter fit.  The reduced $\chi^2$ for the fit is 1.12.  This
fit for the luminosity function is shown as a dashed curve in
Figure~\ref{fig:lf}.

A possible explanation for the flat faint--end slope is that we have
overestimated our completeness at the faint end, by failing to
identify the AGN signatures in faint QSOs, perhaps because faint AGN
might be overwhelmed by the light from their host galaxies.  We do not
believe this to be the case, for several reasons. First, in no case
have we observed ``intermediate'' cases of star forming galaxies with
broad emission lines superposed.  In contrast to virtually all LBG
spectra \citep{shapley2003}, we do not see interstellar absorption
lines in any of the spectra of broad-lined objects at $z \sim 3$.
However, perhaps the strongest argument comes from examining the
spectral properties of identified $z > 2.5$ X-ray sources in the 2~Ms
catalog for the \textit{Chandra} Deep Field North and other very deep
X-ray surveys.  If it were common for AGN to be overwhelmed by their
host galaxies, there would be a significant number of faint X-ray
sources with spectra that resemble those of ordinary star forming
galaxies. To date, virtually all published spectra of objects
identified in the redshift range of interest have obvious AGN
signatures in their spectra, whether they are broad-lined or
narrow-lined AGN.

We can also directly compare the \citet{barger03} CDF--N catalog with
our own color-selected catalog in their region of overlap, with the
following results: There are 2 AGN (1~narrow-lined, 1~QSO) which are
detected by \textit{Chandra} and also discovered in the LBG survey;
there are 2~QSOs which are detected by \textit{Chandra} but did not
have LBG colors in our survey (one of which does have LBG colors in
more recent photometry); and there is 1~QSO detected by \textit
{Chandra} which has LBG colors but is slightly too faint for inclusion
in our survey.  These results are consistent with our overall
completeness estimates, which are approximately 50\% over the range of
redshifts considered.
In addition, there is a narrow-lined AGN (``HDF--oMD49'') discovered
in the LBG survey which is detected in the \textit{Chandra}
exposure~\citep{ccs-agn} at a level below the limit for inclusion in
the main catalog~\citep{A03}; another identified narrow-lined AGN at
$z=2.445$ has no \textit{Chandra}-detected counterpart.  Of the
84~objects with $2.5 \le z \le 3.5$ in our current color-selected
spectroscopic sample in the GOODS--N field, only 2 are detected in the
2~Ms \textit{Chandra} catalog, and both are obvious broad-lined QSOs.

Taken together, all of these arguments suggest that optically faint
QSOs are unlikely to be missed because of confusion with the UV
luminosity of their host galaxies, and thus we believe that our
statistics at the faint end are robust.


\section{Implications for the UV radiation field at $z=3$}
\label{sec:implications}

Having measured the luminosity function of QSOs at $z=3$, we can now
place constraints on their contribution to the UV radiation field at
that redshift.  Integrating over the above parametric fit for the QSO
luminosity function, in its entirety, yields a specific luminosity
density $\epsilon_{1450} = 1.5\times10^{25}~{\rm
erg~s^{-1}~Hz^{-1}~}h~{\rm Mpc^{-3}}$.  The luminosity density from
our parametric fit is 50\% of that predicted from the \citet{pei95}
fit ($\beta_l=1.64$, $\beta_h=3.52$), and is $\sim 8$\% of the UV
luminosity density produced by LBGs at the same redshift based on the
luminosity function of \citet{adelberger00}.

Scaling the results of \citet{hm96}, this fit for the LF produces an
\ion{H}{1} photoionization rate of $\Gamma_{\rm H~I} \approx 8.0
\times 10^{-13}~{\rm s^{-1}}$, which can account for a metagalactic
flux of $J_{912} \approx 2.4 \times 10^{-22}~{\rm
erg~s^{-1}~cm^{-2}~Hz^{-1}~sr^{-1}}$.  The ionizing background
spectrum and \ion{He}{2} ionization fraction, which affect this
calculation, are the results of models and are discussed in detail by
\citet{hm96}.  This value should be considered an upper limit, because
the ability of ionizing photons produced by faint AGN to escape their
host galaxies has not been measured, and may be lower than for the
bright QSOs for which self-absorption in the Lyman continuum is rare
(see SSB)\footnote{A propensity to have a higher fraction of Lyman
continuum ``self-absorbed'' QSOs at faint UV luminosities would
translate directly into an over-estimate of the completeness
correction for a color selected survey such as ours, since an
optically thick Lyman limit at the emission redshift makes a QSO more
likely to be selected using our color criteria and we are missing
fewer than our estimates above would indicate .  In this case, we have
over-corrected the space density, and the there would be even fewer
faint QSOs than we measure above.}.
 
The constraints on the ionizing flux from ``proximity effect''
analyses of the Ly$\alpha$ forest are uncertain, but are still
consistent with the integrated QSO value $z
\sim 3$ \citep[e.g.][]{scott02}. In any case, the ratio of the total
non-ionizing UV luminosity density of star forming galaxies relative
to that of QSOs at $z \sim 3$ implies that QSOs must have a
luminosity-weighted Lyman continuum escape fraction that is $\gtrsim
10$ times higher than that of galaxies if they are to dominate the
ionizing photon budget.

The detection of the \ion{He}{2} Gunn--Peterson effect in $z \sim 3$
QSO spectra has demonstrated that helium reionization occurs during
this epoch.  Unlike the reionization of hydrogen, which can be
effected by radiation from both massive stars and QSOs, the
reionization of helium requires the hard UV radiation produced only by
QSOs.  The improved measurement of the faint end slope, and hence the
total UV luminosity density, will improve simulations of the progress
of reionization, which have previously assumed the \citet{pei95} value
for the faint-end slope~\citep[e.g.][]{sokasian02}.  \citet{miralda00}
have shown that the previously observed bright end of the luminosity
function is sufficient to reionize helium by $z=3$ under most
reasonable assumptions, so the flatter faint-end slope should not
dramatically alter the current picture of \ion{He}{2} reionization;
however, our results may have an effect on the ``patchiness'' of the
reionization as it progresses.

A fortunate consequence of the flat faint-end slope, with implications
for IGM simulations, is that the integrated luminosity $\int
\Phi(L)~L~dL$ converges more rapidly, making the intregral insensitive
to the lower limit of integration.  Simulations will therefore be more
robust, with less dependence on the poorly-understood low-luminosity
AGN population at high redshift.

\section{Conclusions}

Using the 11 QSOs discovered in the survey for $z \sim 3$ Lyman-break
galaxies, we have measured the faint end of the $z=3$ QSO luminosity
function.  This represents the first direct measurement of the faint
end at high redshift.  While the entire luminosity function remains
well-fit by a double power law, the faint end slope differs
significantly from the low-redshift value of $\beta_l=1.64$, being
best fit by a slope $\beta_l=1.24\pm0.07$.  This results in only half
the total QSO UV luminosity compared to previous predictions.  As
measurements of $J_{912}$ from the Ly$\alpha$ forest continue to
improve, we may find that this diminished luminosity from QSOs
requires a substantial contribution from star-forming galaxies.

We believe that the survey described here is successful at detecting
the same broad-lined QSOs that could be detected in even the deepest
X-ray surveys. While the faint X-ray sources that remain unidentified
in the \textit{Chandra} Deep Fields may be heavily obscured AGN of
similar bolometric luminosity at similar redshifts, these objects do
not contribute significantly to the UV luminosity density of the $z
\sim 3$ universe. For the first time, we have measured the space
density of $z \sim 3$ QSOs down to luminosities that account for
essentially all of the UV photon production from AGN.  This
measurement is of primary interest for an understanding of the
physical state of the IGM at high redshift, and not necessarily the
evolution of black hole accretion, which is more difficult to quantify
without extensive multi-wavelength campaigns. Nevertheless, the
results of this paper can be compared directly with a vast literature
observing UV-selected broad-lined AGN. Our results on the QSO
luminosity function suggest that either the mass function and
accretion efficiency of super-massive black holes at $z \sim 3$ is
very different from that at lower redshift, or there has been
significant differential evolution of AGN obscuration as a function of
bolometric luminosity and/or redshift.  If the results are interpreted
as a difference in the mass function of supermassive black holes, then
they may be consistent with some theoretical work which predicts that
low-mass SMBHs form at smaller redshifts than the most massive black
holes \citep[e.g.][]{small92}.

\acknowledgements

We wish to extend special thanks to those of Hawaiian ancestry on
whose sacred mountain we are privileged to be guests.  Without their
generous hospitality, most of the observations presented herein would
not have been possible.  We would also like to thank the staffs at the
Keck and Palomar observatories for their invaluable assistance with
the observations, and the anonymous referee for helpful suggestions.
MPH, CCS, and AES have been supported by grant AST-0070773 from the
U.S. National Science Foundation and by the David and Lucile Packard
Foundation.

\bibliographystyle{apj}

\end{document}